\documentclass{aa}

\usepackage{graphicx}
\usepackage{natbib}
\usepackage{txfonts}

\begin{document}

\title {The Origin of the Mass-Metallicity relation: an analytical approach }

\author {E. Spitoni\inst{1}
\thanks {email to: spitoni@oats.inaf.it}
  \and F. Calura\inst{2}
 \and F. Matteucci\inst{1,2} 
  \and S. Recchi \inst{3}}

\institute{Dipartimento di Fisica, Sezione di Astronomia, Universit\`a di Trieste, via G.B. Tiepolo 11, I-34131  
\and  I.N.A.F. Osservatorio Astronomico di Trieste, via G.B. Tiepolo 11, I-34131
\and Institute of Astronomy,
  Vienna University, T\"urkenschanzstrasse 17, A-1180, Vienna,
  Austria }

\date{Received xxxx / Accepted xxxx}

\abstract {The existence of a mass-metallicity (MZ) relation in star
  forming galaxies at all redshift has been recently established. It
  is therefore important to understand the physical mechanisms
  underlying such a relation.}  {We aim at studying some possible
  physical mechanisms contributing to the MZ relation by adopting
  analytical solutions of chemical evolution models including infall
  and outflow.}  {Analytical models assume the instantaneous recycling
  approximation which is still an acceptable assumption for elements
  produced on short timescales such as oxygen, which is the measured
  abundance in the MZ relation.We explore the hypotheses of a variable
  galactic wind rate, infall rate and yield per stellar generation
  (i.e. a variation in the IMF), as possible causes for the MZ
  relation.}  {By means of analytical models we compute the expected O
  abundance for galaxies of a given total baryonic mass and gas
  mass.The stellar mass is derived observationally and the gas mass is
  derived by inverting the Kennicutt law of star formation, once the
  star formation rate is known. Then we test how the parameters
  describing the outflow, infall and IMF should vary to reproduce the
  MZ relation, and we exclude the cases where such a variation leads
  to unrealistic situations. } {We find that a galactic wind rate
  increasing with decreasing galactic mass or a variable IMF are both
  viable solutions for the MZ relation. A variable infall rate instead
  is not acceptable.  It is difficult to disentangle among the outflow
  and IMF solutions only by considering the MZ relation, and other
  observational constraints should be taken into account to select a
  specific solution. For example, a variable efficiency of star
  formation increasing with galactic mass can also reproduce the MZ
  relation and explain the downsizing in star formation suggested for
  ellipticals. The best solution could be a variable efficiency of
  star formation coupled with galactic winds, which are indeed
  observed in low mass galaxies.}


\titlerunning{ }
\authorrunning{Spitoni et al.}
\maketitle

\section{Introduction}

The study of the relation between the galactic stellar mass and the
gas-phase metallicity, which represents the abundance of heavy
elements present in the interstellar medium (hereinafter ISM),
provides us with constraints on various parameters fundamental for any
galaxy formation theory.  These parameters include the stellar initial
mass function (IMF), representing the distribution of stellar masses
at birth, which strongly influences the amount of heavy elements
produced and restored by stars into the ISM.  The mass-metallicity
(hereinafter MZ) relation can also provide information on the star
formation history: in general, a higher star formation rate produces a
larger concentration of metals in the ISM.  Finally, the MZ relation
provides with information on the role of infall/outflow in galaxy
evolution, namely on the amount of matter either ejected from galaxies
into the inter galactic medium (IGM) or accreted onto galaxies from
the IGM. Galactic outflows are directly connected to the importance of
feedback and energy exchange between stars and interstellar gas.  At
the present time, different theoretical explanations of the MZ
relation have been proposed by various authors.  The first explanation
is based on starburst-induced galactic outflows, more efficient in
expelling metal-enriched matter from low-mass galaxies than from giant
galaxies, mainly owing to the shallower gravitational potential wells
of the former (Larson 1974; Dekel \& Silk 1986; Tremonti et al. 2004;
De Lucia et al. 2004; Kobayashi et al. 2007; Finlator \& Dav\'e 2008).
However, at the present time it is difficult to assess how the outflow
efficiency depends on the baryonic galactic mass.  In an alternative
scenario, infall of pristine gas can dilute the interstellar metals
and act in the same way as outflows, once one assumes longer infall
time scales in lower mass galaxies (Dalcanton et al. 2004, but see
also Dalcanton 2007).  Another explanation is that dwarf galaxies are
less evolved than large galaxies, namely that the efficiency of star
formation is larger in more massive systems.  In this picture, large
galaxies have formed the bulk of their stars by means of an intense
star formation event at high redshift, quickly enriching their ISM to
solar or over-solar metallicities, whereas dwarf galaxies,
characterized by lower star formation efficiencies (i.e. star
formation rates per unit mass of gas) have sub-solar interstellar
metallicities. This interpretation is supported by various chemical
evolution studies (Lequeux et al. 1979; Matteucci 1994; Calura et
al. 2009), by cosmological N-body simulations (Brooks et al. 2007;
Mouhcine et al. 2008; Tassis et al. 2008) and by hydrodynamical
simulations (Tissera et al.  2005; De Rossi et al. 2007).  A fourth
way to produce a MZ relation is by means of a variable IMF.  K\" oppen
et al. (2007) showed how a higher upper mass cutoff and a flatter
slope in the IMF in more massive galaxies can account for the observed
MZ relation.  It is possible that the relative importance of these
processes may vary as a function of the total galactic stellar
mass. However, the theoretical investigations performed so far do not
allow us to disentangle among these phenomena and to assess which may
be considered dominant in galaxies of different mass.

In this paper, we aim at investigating the importance of different
physical processes in determining the MZ relation, by means of
analytical chemical evolution models.  This analytical approach will
allows us to rapidly test the parameter space involved in our study,
and it is justified by the fact that the oxygen evolution can be
computed under the instantaneous recycling approximation, required to
obtain analytical solutions.  By means of the solutions of the
analytical models including infall and outflow we can compute the
expected metallicity for a galaxy of a given mass and a given gas mass
fraction. The total stellar mass and gas mass are derived from the
observational data: in particular, we use the local MZ relation
determined in SDSS galaxies, and from the SFRs observed in the same
galaxy sample, we derive the gas mass fractions by adopting the
Kennicutt (1998) law for star formation.  We aim at assessing the
importance of outflow, infall and IMF (through the yield per stellar
generation) in determining the MZ relation for galaxies of various
masses.  We also aim at providing detailed expressions for the
effective yields in different physical conditions, that can be used to
compare with the empirical effective yields as derived from
observations (see Dalcanton 2007).  \\

Our paper is organized as follows. In section 2, we present our
equations, our basic assumptions and the observables used in our
study. In Section 3, we present our results and finally, in Section 4,
some conclusions are drawn.

\section{Summary of simple model solutions}
\label{simple}

\subsection{The Simple model}
\label{sec:standard}
 We recall here the main assumptions and results of the simple models of
chemical evolution (Tinsley 1980; Matteucci 2001; Recchi et al. 2008).
As it is well known, the so-called {\it Simple Model} of chemical
evolution is based on the following assumptions:

\begin{enumerate}

\item the system is one-zone and closed, namely there are no inflows 
or outflows.

\item The initial gas is primordial (no metals).

\item The IMF is constant in time.

\item The gas is well mixed at any time ({\it instantaneous mixing approximation}).

\item Stars more massive than 1 M$_\odot$ die instantaneously; stars 
smaller than 1 M$_\odot$ live forever ({\it instantaneous recycling
approximation} or IRA).

\end{enumerate}
\noindent
These simplifying assumptions allow us to calculate analytically the
chemical evolution of the galaxies.  Once we have defined the
fundamental quantities, such as the returned fraction:

\begin{equation}
R = \int_1^\infty (m - M_R) \phi (m) dm,
\label{eq:r}
\end{equation}
\noindent
(where $\phi (m)$ is the IMF and $M_R$ is the mass of the remnant) and the yield per 
stellar generation:

\begin{equation}
y_Z = {1 \over {1 - R}} \int_1^\infty m p_{Z, m} \phi (m) dm,
\label{eq:yield}
\end{equation}
\noindent
(where $p_{Z, m}$ is the fraction of newly produced and ejected metals
by a star of mass $m$)

The well known solution of the closed box
model can be easily found:

\begin{equation}
Z = y_Z \ln (\mu^{-1})
\label{eq:simple}
\end{equation}

\noindent
where $\mu$ is the gas fraction $M_{gas}/M_{tot}$, with $M_{tot}=M_* +
M_{gas}$.  This result is obtained by assuming that the galaxy
initially contains only gas and has the remarkable property that it
does not depend on the particular star formation history the galaxy
experiences. Moreover, to obtain eq.  (\ref{eq:simple}) one has to
assume that $y_Z$ is constant in time.  The yield which appears in
eq. (\ref{eq:simple}) is known as effective yield, $y_{Z_{eff}}$,
namely the yield the system would have if described by the simple
model. In models with gas flows the true yield, as defined in eq. (2),
does not coincide with the effective yield, as we will see in the next
paragraphs.

\subsection{Leaky box models}       
Relaxing the first of the assumptions of the simple model, we get the
models including gas flows, also known as {\it leaky box} models.
Analytical solutions of simple models of chemical evolution including
infall or outflow are known since at least 30 years (Pagel \& Patchett 1975
; Hartwick 1976 ; Twarog 1980; Edmunds 1990
).  Here we follow the approach and the terminology of
Matteucci (2001), namely we assume
for simplicity {\it linear} flows (we assume gas flows proportional to
the star formation rate (SFR)).  Therefore, the outflow rate $W (t)$
is defined as:
 
\begin{equation}
W (t) = \lambda (1 - R) \psi (t),
\label{eq:w}
\end{equation}
\noindent
where $\psi (t)$ is the SFR, and the infall rate $A(t)$ is given by:

\begin{equation}
A (t) = \Lambda (1 - R) \psi (t).
\label{eq:a}
\end{equation} 
\noindent
Here $\lambda$ and $\Lambda$ are two proportionality constants $\ge
0$.  The first assumption is justified by the fact that, the larger
the SFR is, the more intense are the energetic events associated with
it (in particular supernova explosions and stellar winds)and therefore
the larger is the chance of having a large-scale outflow (see
e.g. Silk 2003).  A proportionality between $A (t)$ and $\psi (t)$ is
 less easily justifiable physically.

If we consider a system including outflow and infall the set of equations we need to solve is the following one:

\begin{equation}
\cases{{d M_{tot} \over d t} = (\Lambda - \lambda) (1 - R) \psi (t) \cr
{d M_{gas} \over d t} = (\Lambda - \lambda - 1) (1 - R) \psi (t) \cr
{d M_Z \over d t} = (1 - R) \psi (t) [\Lambda Z_A + y_Z - (\lambda + 1) Z]}
\label{eq:system}
\end{equation}
\noindent
where $M_Z$ is the mass of metals ($M_Z=Z \cdot M_{gas}$) and $Z_A$ is the
metallicity of the infalling gas.

\subsubsection {Simple Model with outflow} 
If we study a simple model  with  only outflow, e.g. $A(t)=0$,  the system of equation (\ref{eq:system}) can be worked out, yielding to the following solution (Matteucci 2001): 

\begin{equation}
Z = {{y_Z} \over (1+\lambda)}\ln [(1+\lambda)\mu^{-1}-\lambda].
\label{zoutflow}
\end{equation}
 \noindent
 
For the integration we assume that at $t=0$, $Z$(0)=0,
$M_{tot}(0)=M_{gas}(0)=M_{g,0}$. Since $\mu$ values span in the range
between $0 \le \mu \le 1$, the solution (\ref{zoutflow}) is well
defined for all the values of $\lambda \ge 0$.
In this case, the true yield is given by:

\begin{equation}
y_Z = {Z (1+\lambda) \over \ln [(1+\lambda)\mu^{-1}-\lambda]}
\end{equation}

and it is clearly lower than the {\it effective yield} defined in eq. (3). This fact should be kept in mind when using eq. (3) to interpret observational data.

\subsubsection {Simple Model with infall}
Here we analyze the opposite case: $A(t) \ne 0 $ and $W(t)=0$. 
Assuming for the infalling gas a primordial composition  (e.g $Z_A=0$), $\Lambda \ne 1$ and  $\Lambda \ne 0$, we obtain this solution (Matteucci 2001):
\begin{equation}
Z = {{ y_Z} \over \Lambda}\biggl\lbrace 
1 - \bigl[ (\Lambda- (\Lambda  - 1) \mu^{-1} \bigr]^{\Lambda \over {\Lambda   
- 1}} \biggr\rbrace.
\label{zinfall}
\end{equation}
\noindent
We notice here  that, in models with infall, not all the values of $\mu$ are
allowed.If $\Lambda >   1$,   $\mu$ ranges between 1 and
a minimum value:

\begin{equation}
\mu_{\rm min} = {{\Lambda - 1} \over {\Lambda}}.
\label{eq:mumin}
\end{equation}
\noindent
In the case with $\Lambda <1$ there is no $\mu_{min}$. If $\Lambda =
1$ the solution is: $Z=y_z[1 -e^{-(\mu^{-1}-1)}]$, which is well known
solution of the extreme infall (Larson 1972), where the amount of gas
remains constant in time.
\subsubsection {Simple Model with inflow and outflow}
The general solution for a system described by the simple model in
the presence of infall of gas with a general metallicity $Z_A$ and
outflow is (Recchi et al. 2008):
\begin{equation}
Z = {{\Lambda Z_A + y_Z} \over \Lambda}\biggl\lbrace 
1 - \bigl[ (\Lambda - \lambda) 
- (\Lambda - \lambda - 1) \mu^{-1} \bigr]^{\Lambda \over {\Lambda - \lambda 
- 1}} \biggr\rbrace
\label{eq:sol}
\end{equation}
\noindent
If the outflow rate is larger than the infall rate the solution is defined for all $\mu$ values, in the opposite case not all values of $\mu$ are allowed.

In the models in which $\Lambda > \lambda + 1$, $M_{gas}$ is always
increasing (eq. \ref{eq:system}), therefore $\mu$ ranges between 1 and
a minimum value:

\begin{equation}
\mu_{\rm min} = {{\Lambda - \lambda - 1} \over {\Lambda - \lambda}}.
\label{eq:mumin}
\end{equation}
\noindent

For models in which $\lambda + 1 > \Lambda > \lambda$ there is no
$\mu_{\rm min}$ but there is a upper limit reachable by the gas mass
which is given by $M_{gas, lim} = M_{g, 0} / (\lambda + 1 - \Lambda)$.
\subsubsection {Simple Model with differential winds}

Recchi et al. (2008) presented a set of new solutions in the simple
model context in presence of differential winds, namely models where
the metals are more easily channelled out of the parent galaxy than
the pristine gas.  The easiest way to consider a differential wind in
the framework of simple models of chemical evolution is to assume that
the metallicity of the gas carried out in the galactic wind is
proportional to the metallicity of the ISM with a proportionality
constant larger than one.  If we define $Z^o$ as the metallicity of
the outflowing gas, this condition implies that $Z^o = \beta Z$ with
the ejection efficiency $\beta > 1$.  In the metallicity budget
(third equation in (\ref{eq:system})) we assume that the negative term
due to the galactic wind is given by $W (t) Z^o = \beta \lambda (1 -
R) \psi (t)$.

With their simple approach, Recchi et al. (2008)  were able to
determine analytical expressions for the evolution of $Z$, which allow
us to understand more clearly the effect of galactic winds on the
chemical evolution of galaxies.

The set of the equations we solve in this case is very similar
to (\ref{eq:system}), with the only difference given by the
metallicity budget equation, which we modify as follows:

\begin{equation}
{d M_Z \over d t} = (1 - R) \psi (t) [\Lambda Z_A + y_Z - (\lambda \beta 
+ 1) Z].
\label{eq:diffw}
\end{equation}
\noindent
The solution of this new set of equations is given by:

\begin{equation}
Z = {{\Lambda Z_A + y_Z} \over {\Lambda + (\beta - 1)} \lambda}\biggl\lbrace 
1 - \bigl[ (\Lambda - \lambda) 
- (\Lambda - \lambda - 1) \mu^{-1} \bigr]^
{{\Lambda + (\beta - 1) \lambda} \over {\Lambda - \lambda - 1}} 
\biggr\rbrace.
\label{eq:diffwsol}
\end{equation}
\noindent
It is trivial to see that we can obtain eq. (\ref{eq:sol}) in the case $\beta$=1 (i.e. in the case in which the galactic wind is not differential).

\section{The Observed  Mass-metallicity relation and our method}

\begin{figure}
\includegraphics[width=0.45\textwidth]{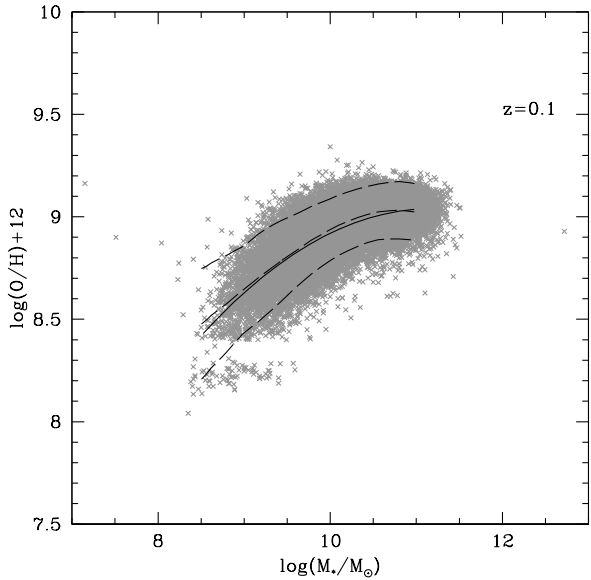}
\caption{The observed MZ relation for the oxygen (Kewley \& Ellison,
  2008). The dashed lines are the average values and the standard
  deviation and the solid one is the best fit to the average values as
  given by Maiolino et al. (2008).The redshift of the galaxies is
  indicated at the top right of the figure.}
\label{massmet}
\end{figure}

 Kewley \& Ellison (2008) analyzed the metallicity relation for Sloan
 Digit Sky Survey (SDSS) 27,730 star-forming galaxies by adopting 10
 different metallicity calibrations. In Fig. \ref{massmet} we show the
 MZ relation at $z=0.1$ obtained by using the calibration of Kewley \&
 Dopita (2002), considered the best in the analysis by Calura et
 al. (2009), including the average values and the standard deviation
 indicated with the dashed lines.

 In the same figure the  solid line is the analytical fit of the observed MZ relation of  Kewley \& Ellison (2008), as  given by Maiolino et al. (2008), in particular:  
\begin{equation}
\log(O/H) +12=-0.0864*\left[\log\left( \frac{M_{*}}{M_{\odot}}\right) -11.18\right]^{2}+9.04.
\label{maio}
\end{equation}
\noindent
\begin{figure}
\includegraphics[width=0.45\textwidth]{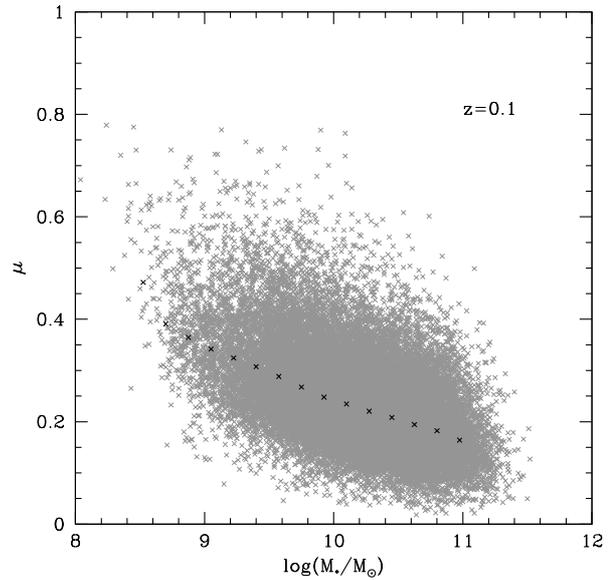}
\caption{The gas fraction $\mu$ as a function of the stellar galactic mass $M_*$. We divide the stellar masses in bins and  the average  $\mu$ values for each bin are indicated by black crosses.}
\label{mu}
\end{figure} 
\noindent
Since now on we will use this analylitical fit as our fiducial MZ relation. For  estimating  the amount of gas which resides in each star-forming galaxy we use the method described in  Calura et al. (2008). We aim at determining the cold gas mass of each galaxy on the basis of its
SFR, by inverting the Kennicutt (1998) (hereafter K98) relation, which links the
gas surface density to the SFR per unit area.
A similar technique was used by  Erb et al. (2006) and Erb (2008) to derive the gas fractions for  a sample of star forming 
galaxies at $z\sim 2$ and to study the implications of the MZ relation observed at high 
redshift on the galactic gas accretion history, respectively. \\
Following K98, for any galaxy,
the gas surface density $\Sigma_{gas}$, expressed in $M_{\odot} \, pc^{-2}$, depends on the SFR surface
density $\dot{\Sigma_*}$, expressed in $M_{\odot} \, yr^{-1} \, kpc^{-2}$ according to:

\begin{equation}
 \Sigma_{gas} = \big( \frac{ \dot{\Sigma_{*}} }{ 2.5 \times 10^{-4}}\big)^{0.714} \, \, \, \, \,    M_{\odot} \, pc^{-2}
\end{equation}

 The  scaling radius $R_d$, which will be used to compute the SFR surface density profile of each galaxy, can be calculated as (Mo et al. 1998):
\begin{equation}
R_d=\frac{\lambda_s  R_{200}f_{c_{vir}}^{-1/2}f_{R}(\lambda_s,c_{vir},f_{b})}{\sqrt{2}}.
\label{6}
\end{equation}
$\lambda_s$ is the spin parameter of the halo and depends on the total energy of the halo $E$,
its angular momentum $J$ and its mass $M$ according to:
\begin{equation}
\lambda_S=J|E|^{1/2}G^{-1}M^{-5/2}.
\label{8}
\end{equation}
The quantity $\lambda_s$ is likely to assume values in the range $0.01\le \lambda\le 0.1$ (Heavens \& Peacock (1988), Barnes \& Efstathiou (1987), Jimenez et al. 1998).
In this paper, we assume a typical value of $\lambda_s=0.05$.  Scatter in $\lambda_s$ would propagate as scatter in the gas fractions, but we average over many galaxies to obtain the average gas fractions.\\
The parameter $c_{vir}$ is the halo concentration factor, and is calculated following Bullock et al. (2001) and Somerville et al. (2006) i.e.,
by defining for each halo a collapse redshift $z_c$, as $M_{*}(z_c)=F \cdot M$.
$c_{vir}$  is given by $c_{vir}(M,z)=K(1+z_c)/(1+z)$, where $F$ and $K$ are two adjustable parameters.
Following Somerville et al. (2006), we assume $K=3.4$ and $F=0.01$. Here we assume $z\sim 0.1$, corresponding to the average redshift of the SDSS galaxies (Tremonti et al. 2004).  \\
To compute the quantities $f_{c_{vir}}$ and $f_{R}$, it was used an analytic fitting functions presented in Mo et al. (1998). 
For each galaxy, if $\psi$ is the SFR in units of $M_{\odot}/yr$ and if we assume for the
SFR surface density profile $\dot{\Sigma}_{*}(R)=\dot{\Sigma}_{*,0} \, \exp (-1.4 R/R_{d})$,
the central SFR surface density is then given by:
\begin{equation}
\dot{\Sigma}_{*,0} = \frac{\psi}{2 \pi (R_{d}/1.4)^{2}}.
\end{equation}
The gas surface density $\Sigma_{gas}$ is then determined by the K98 relation, and the gas mass $M_{gas}$
(in $M_{\odot}$) is given by:
\begin{equation}
 M_{gas} = \Sigma_{gas} \times 2 \pi  R_{d}^{2}.
\end{equation}
At this point we have a relation between  $\mu$ and  the stellar mass  for each considered galaxy as shown in Fig. \ref{mu}. 

\begin{figure}
\includegraphics[width=0.45\textwidth]{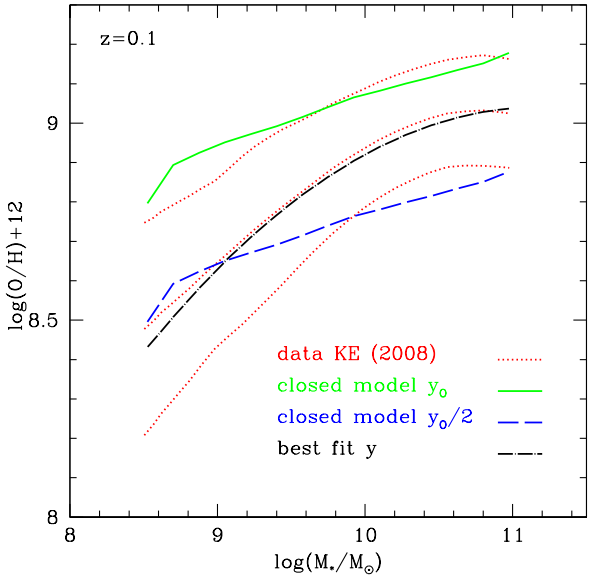}
\caption{The observed MZ relation and  related standard deviation are indicated with the dotted red lines. With the green solid line we show our closed box model results assuming for the oxygen a yield per stellar generation of $y_O=0.01$, whereas with the dashed blue line we show the model with $y_O=0.005$. The model which best fits the data by using a variable yield is labelled with the short dashed dotted line.}
\label{yfit}
\end{figure} 

\begin{figure}
\includegraphics[width=0.45\textwidth]{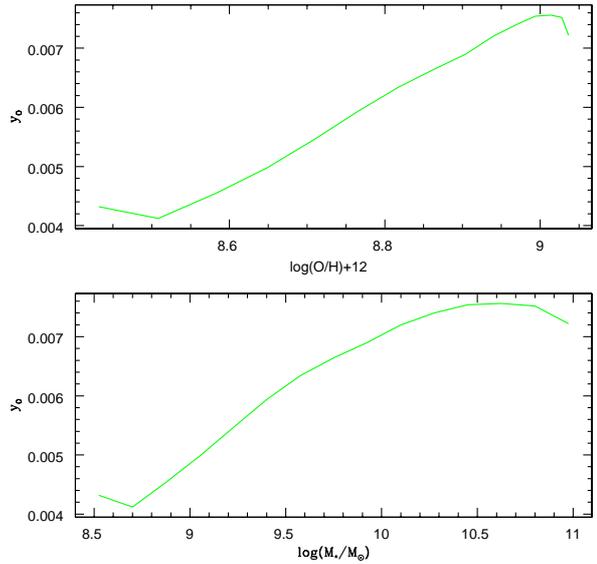}
\caption{In the lower panel we report the $y_O$ yield values as functions of the galactic stellar mass $M_*$, those obtained to reproduce  the MZ relation, while in the upper panel we show the same yields as functions of the derived O abundances.}
\label{y}
\end{figure}

The aim of this work is to constrain the model parameters of  simple models to reproduce the analytical fit (eq. \ref{maio}) of the MZ relation. In our reference model we assume for the oxygen yield the value of $y_O=0.01$, this yield is obtained by adopting the Salpeter IMF (x=1.35 over the mass range 0.1-100$M_{\odot}$), the same used for the inversion of the K98 law, and the stellar yields of Woosley \& Weaver (1995) for oxygen at solar metallicity.  In the first part of our work we explore the effect of a different choice of  yields in the  closed models. Given a stellar mass content $M_{*}$, we have an oxygen abundance value from the observed MZ relation, and  then for each galaxy  we have an estimate of the gas fraction $\mu$ using the procedure described above. Therefore in closed models the only free parameter is the yield. We will show also the effect of a non-constant yield,
assuming that is the IMF which varies and not the nucleosynthesis from galaxy to galaxy.

Concerning  leaky box models if we fix  $Z$ and $\mu$ we can vary the wind and infall parameters as well as the yield. We aim at testing the way in which the parameter space  is constrained if we want to reproduce the observed MZ relation using simple model solutions.

\begin{figure}
\includegraphics[width=0.45\textwidth]{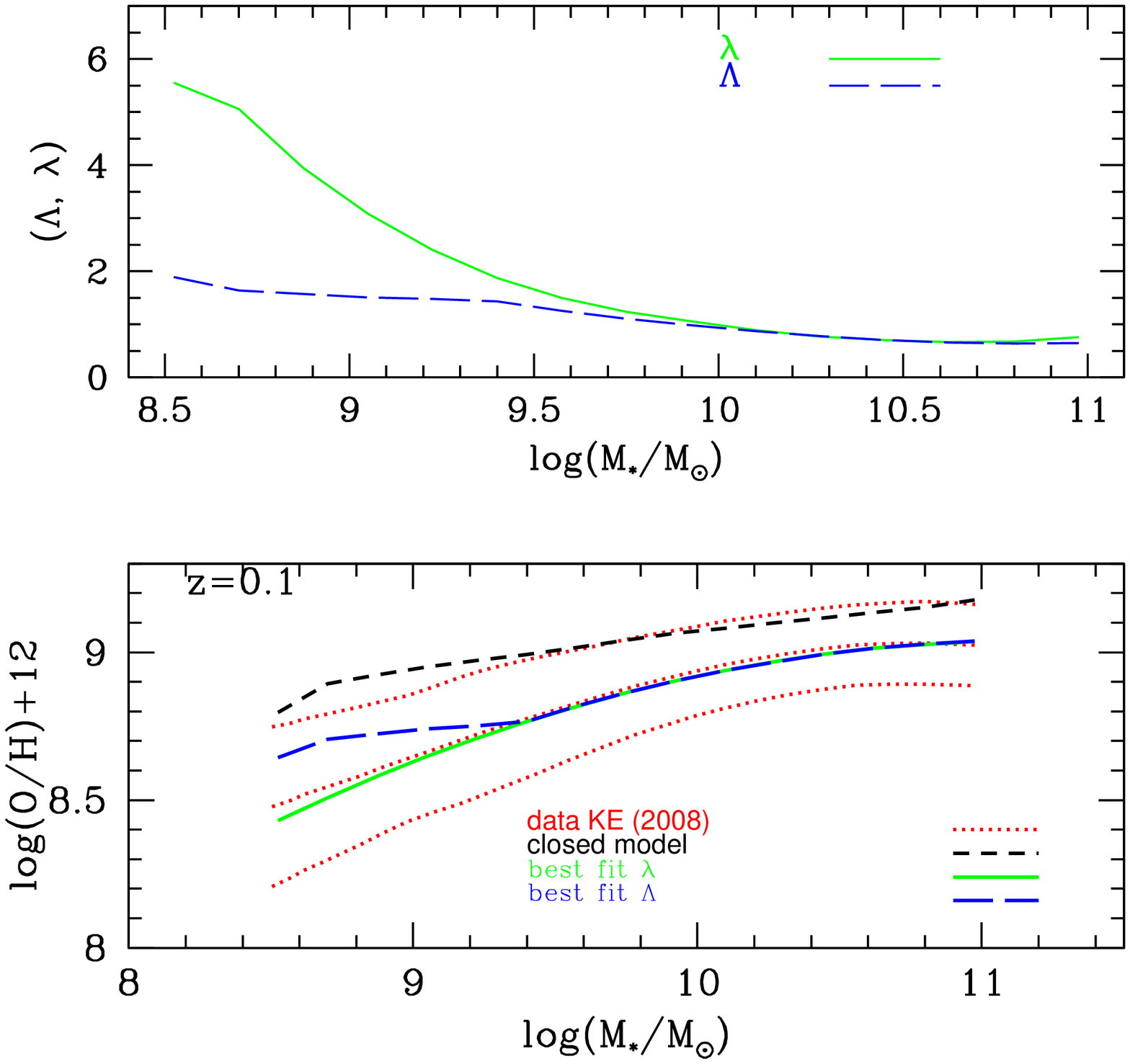}
\caption{ {\it Upper panel}: The best $\Lambda$ and $\lambda$ values obtained for models with only infall and outflow, respectively, in order to reproduce the MZ relation. $\Lambda$ and $\lambda$ are given as functions of the galactic stellar mass $M_*$.
  {\it Lower panel}: The best fit to the observed MZ relation and  related standard deviation are indicated with the dotted red lines. With the green solid line we show our best model results considering only outflow, while with the long dashed blue line the best  results for the model with only infall are. With the short dashed line are indicated the results of the closed box model. In all models a value of $y_O=0.01$ is adopted.}
\label{lL}
\end{figure} 

\begin{figure}
\includegraphics[width=0.45\textwidth]{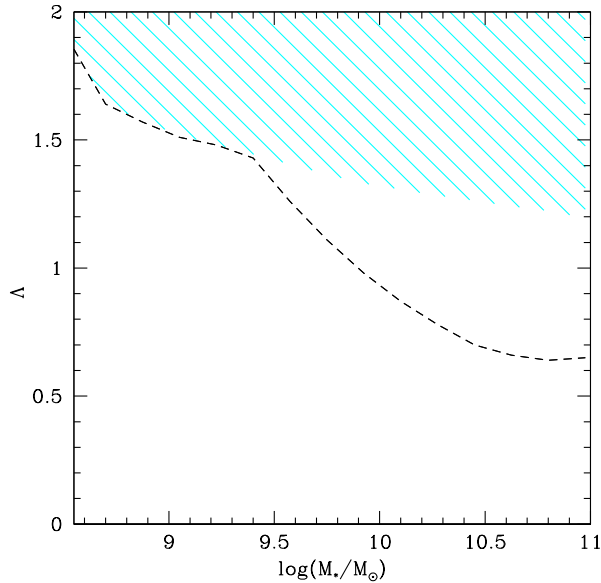}
\caption{ Case with only infall: dashed area represents the forbidden $\Lambda$ values, the rest of the area represents the allowed values. With the short dashed line we draw the values of  $\Lambda$ as a function of $\log \left(M_*/M_{\odot} \right)$ adopted in the MZ relation of Fig. \ref{lL}.  }
\label{area}
\end{figure}

\section{Results}
\subsection{Closed Box Model Results}
First of all we report our results for the closed model. In this case we can vary only the yield in the eq. (\ref{eq:simple}), because once fixed $M_*$, the Z is given by observational fit of Maiolino et al. (2008) and $\mu$ from the inversion of the Kennicutt law. In the Fig (\ref{yfit}) we report the simple model solution compared to  the observational average values and the standard deviation for the MZ relation.
If we consider the simple model adopting a constant oxygen effective yield of $y_O$=0.01,  observational data are clearly not reproduced.  The predicted slope is too flat as shown in Fig (\ref{yfit}) with the green solid line.
Even if we consider a smaller yield,  the problem is not solved: the MZ is simply translated at lower values with the same slope (the blue dashed line in Fig \ref{yfit} is the model with $y_O$=0.005). The next step is to consider a variable yield as a function of the $M_*$. In the upper panel of Fig. \ref{y} we report the  yields per stellar generation as a function of the metallicity  obtained to
reproduce  the MZ relation, whereas  in the lower panel $y$ is as a function of stellar mass. The best fit yields spans in the range between 0.004 and 0.007.

\subsection{Leaky box Results}

In this section we show our results in the Leaky box framework. We keep the oxygen yield fixed at $y_O=0.01$. First of all we present  results for systems in which only outflows (eq. \ref{zoutflow}) and only infall (eq. \ref{zinfall})  are included, separately. In the upper panel of Fig \ref{lL} we report the best fit $\Lambda$ and $\lambda$ parameters as a function of stellar mass in galaxies for reproducing the MZ relation. Both the parameters, as expected,  anti-correlate with the stellar mass in galaxies. We note that  for stellar masses smaller than $ 10^{9.5}$, we have $\lambda >> \Lambda$. If we look at the MZ relation plot obtained with the best fits for $\lambda$ and $\Lambda$ as reported in  the lower panel of Fig. \ref{lL}, we see that for   $ M_* < 10^{9.5}M_{\odot}$ the  model with only infall  does not reproduce at all the observed MZ relation.
The reason for this is  the fact that for values of $\Lambda>1$ the condition (10) holds, and
this relation can be seen as a condition for $\Lambda$ values:
\begin{equation}
\Lambda \le \frac{1}{1-\mu},
\label{mucond}
\end{equation}
\noindent
\begin{figure}

\includegraphics[width=0.45\textwidth]{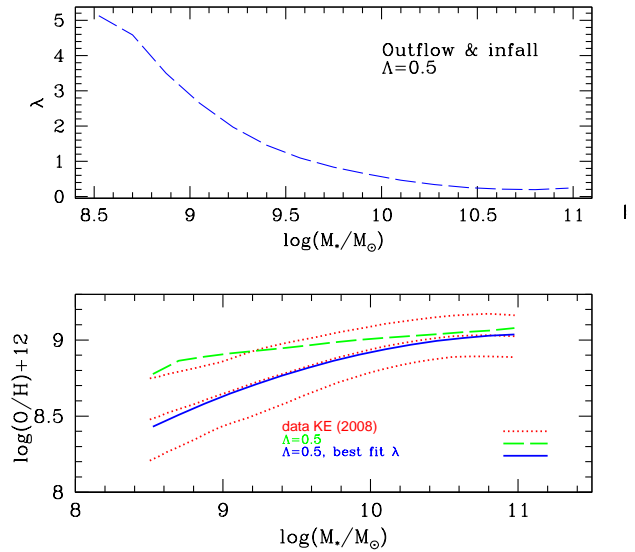}
\caption{{\it Upper panel}: the best  $\lambda$ values to reproduce the MZ relation in the framework of the model with both infall and outflow. The infall parameter is fixed to be  $\Lambda$=0.5. 
{\it Lower panel}:  the best fit to the observed MZ relation and  related standard deviation are indicated with the dotted red lines. With the blue solid line we show the results of the best model with both outflow and infall, obtained by fixing $\Lambda$=0.5 and varying $\lambda$. With the dashed green line is shown the model results with only infall with a constant $\Lambda=0.5$.}
\label{05}
\end{figure} 

  In Fig. \ref{area}  the dashed area represents the forbidden $\Lambda$ values, the rest of the area represents the allowed values. With the short dashed line we draw the values of  $\Lambda$ as a function of $\log \left(M_*/M_{\odot} \right)$ relative to the predicted MZ relation of Fig. \ref{lL}.

When $\log(M_*/M_{\odot})$=8.5 and $\mu=0.47$, the $\Lambda$ values must be $ < $1.88, whereas for  $\log(M_*/M_{\odot})$=11.0  and $\mu=0.16$ we have that $\Lambda < 1.2$. Therefore, for the lower range of stellar masses considered, the derived $\Lambda$ does not allow to reproduce the observed MZ relation. We conclude that the simple model solution concerning system with only infall varying with galactic stellar mass is unable to fit the MZ for small stellar mass galaxies  even if we choose the best fit $\Lambda$ values.

Concerning the solution of simple models  with infall and outflow at the same time, we  fix $\Lambda=0.5$ in the eq. (\ref{eq:sol}), and we test which values  of  $\lambda$ must be assumed   to reproduce the MZ relation. As reported in the upper panel of Fig. \ref{05},  the best fit model for $\lambda$ anti-correlates with the stellar mass $M_*$.

\begin{figure}
\includegraphics[width=0.45\textwidth]{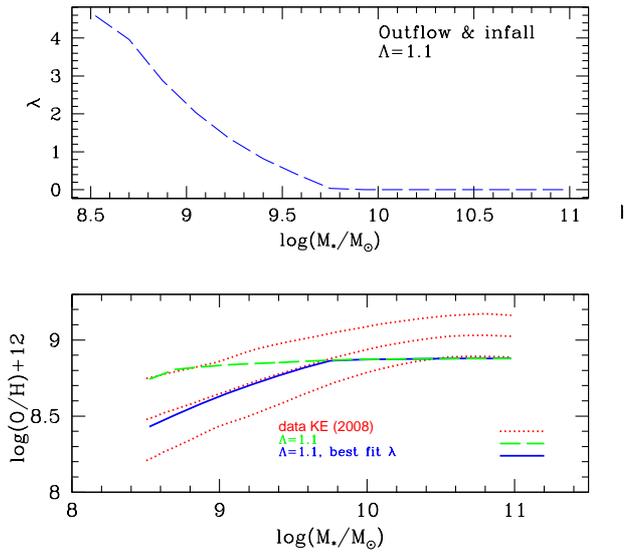}
\caption{{\it Upper panel}:The best  $\lambda$ values for a model with both infall and outflow. $\lambda$ is given as a function of the galactic stellar mass $M_*$ with fixed $\Lambda$=1.1.
{\it Lower panel}: the best fit to the observed MZ relation and  related standard deviation are indicated with the dotted red lines. With the blue solid line we show  the best model results, obtained by considering both outflow in infall and by fixing $\Lambda$=1.1 while varying $\lambda$. With the dashed green line is shown the model solution with only infall with $\Lambda=1.1$.}
\label{1L}
\end{figure}

 In the lower panel we see that the MZ relation is well reproduced by the best fit model line (the blue solid one). In the MZ plot we also include the result of the model only with the infall parameter $\Lambda$  equal to 0.5. This model solution is reported with the dashed line, and clearly it does not reproduce  the MZ relation. Therefore, a variable infall efficiency or a fixed one coupled with a variable outflow as a function of the galactic stellar mass is required.

\begin{figure}
\includegraphics[width=0.45\textwidth]{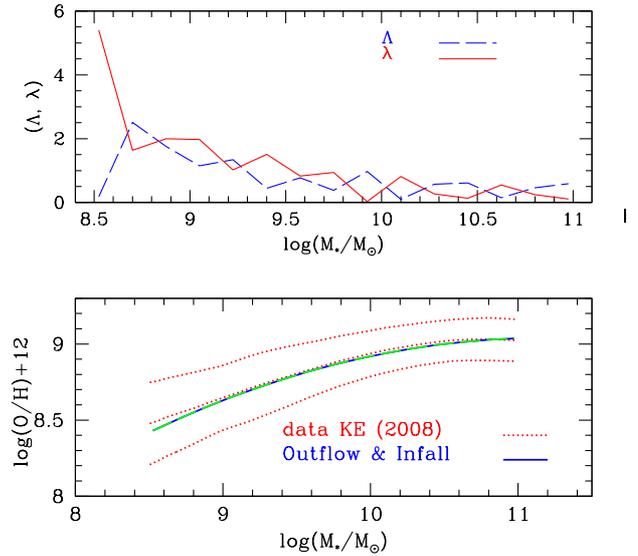}
\caption{{\it Upper panel}: the best  fit values for $\Lambda$ and $\lambda$ parameters in a  model with infall and outflow, varying both the parameters at the same time.
{\it Lower panel}: the bestfit to the observed MZ relation and related standard deviation are indicated with the dotted red lines. With the blue solid line we show the results of the best model  adopting the $\Lambda$ and $\lambda$ values of the upper panel.
With the dashed green line is shown the analytical fit of Maiolino et al. (2008).}
\label{tut}
\end{figure}

We analyze also the case with a constant infall $\Lambda=1.1$ and a variable outflow as reported in Fig. \ref{1L}. In the upper panel we see the behavior of the model with the best fit $\lambda$ values.
The choice of this particular value of $\Lambda$ is just arbitrary but  it shows the effect of a variable galactic wind in presence of an  infall with a fixed $\Lambda$ value, large enough to make the metallicity of large galaxies to decrease. We see that for $M_*$ larger than $10^{9.8} M_{\odot}$,  $\lambda$ vanishes. The reason for this  can be understood analyzing the lower panel of  Fig. \ref{1L}. As in Fig. \ref{05} we also report the solution with only the infall. We note that the infall with $\Lambda=1.1$ is too strong and  flattens the MZ relation for high stellar mass systems  (the green dashed line). Therefore, if we couple this infall with  an variable outflow, the best fit model (blue solid line) is able to fit the MZ relation for systems with  small stellar mass values but certainly we cannot reproduce the part of the observed MZ where the infall was already too strong, namely for high galactic stellar masses.  We find that the maximum value for the $\Lambda$ parameter, when varying $\lambda$ to reproduce the MZ relation, must be $\leq 0.7$.

The results obtained varying both the parameters are reported in Fig. \ref{tut}. In the upper panel, the best fit parameters as a function of $M_*$ are reported. These sets of best parameters lead to a difference in the MZ relation less than $10^{-5}$ dex  between the best fit of Maiolino et al. (2008)  and simple model solutions with this choice. In fact,  in the lower panel of Fig.\ref{tut} comparing the Maiolino's fit and our best fit, we see that  the two lines overlap perfectly. It is worth noting that in most of
the mass bins (at least for $M_*<10^{9.5} M_{\odot})$ is $\lambda>\Lambda$.

\begin{figure}
\includegraphics[width=0.45\textwidth]{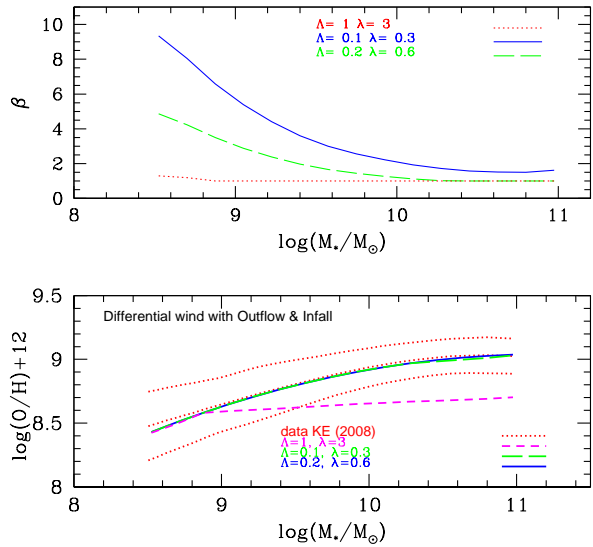}
\caption{{\it Upper panel}: The best  values for $\beta$ in a model with both infall and outflow. The blue solid line is referred to a system with $\Lambda=0.1$ and $\lambda=0.3$, the results for  $\Lambda=0.2$ and $\lambda=0.6$  are indicated with the long 
dashed green line, and  with the red short dashed line is shown the result with $\Lambda=1$ and $\lambda=3$.
{\it Lower panel}: the bestfit to the observed MZ relation and  related standard deviation (the dotted red lines) are shown. 
With the blue solid line we show the results of the best model obtained by varying the ejection efficiency $\beta$ and  including also outflow and infall with $(\Lambda,\lambda)=(0.2,0.6)$. The cases with  $(\Lambda,\lambda)=(0.1,0.3)$ and  $(\Lambda,\lambda)=(1,3)$ are indicated with the green long dashed line and the magenta short dashed line, respectively.}
\label{alfa2}
\end{figure} 

\begin{figure}
\includegraphics[width=0.45\textwidth]{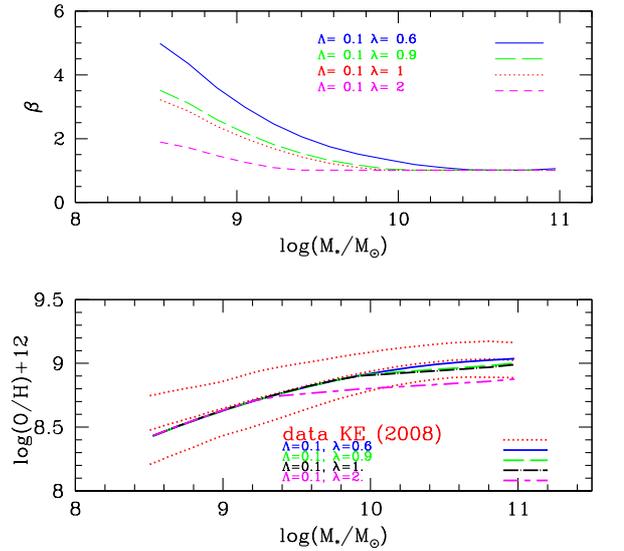}
\caption{{\it Upper panel}: The best  values for $\beta$ in a model with both infall and outflow. We fix here the value of $\Lambda$=0.1 and vary  $\lambda$. 
{\it Lower panel}: the bestfit to the observed MZ relation and related standard deviation are indicated with the dotted red lines. We fix the value of $\Lambda$=0.1. 
The other lines refer to the models of the upper panel.}
\label{alfalimite}
\end{figure}

The last part of our work is focused on simple model solutions with infall and  differential winds. We presented in Section 2, the analytical solution found by Recchi et al. (2008). Considering also in this case a model with primordial infall, i.e. $Z_A=0$ we study systems with 3 sets of parameters for $\Lambda$ and $\lambda$: $(\Lambda,\lambda)=(0.1,0.3);(0.2,0.6),(1,3)$.  In the upper panel of Fig. \ref{alfa2} we show the variation of the possible 
$\beta$ values which can be considered in order to reproduce the MZ relation. As expected, for all models the best fit for $\beta$ anti-correlates with the galactic stellar mass, and the values of $\beta$ are higher for the first model  $(\Lambda,\lambda)=(0.1,0.3)$. Both the first and the second model can reproduce the MZ relation. 
In the case with   $(\Lambda,\lambda)=(1,3)$ instead, we are not able to reproduce the MZ relation, as shown in the lower panel of  Fig. \ref{alfa2} with the magenta short dashed line.
Moreover, as shown in Fig. \ref{alfalimite} we fix the infall parameter at the value $\Lambda=0.1$ and we find that the upper limit value for $\lambda$, in order to fit the MZ relation, is $\sim 0.6$ and in this case  $\beta$ ranges between 1 and 5. 
\begin{figure}
\includegraphics[width=0.45\textwidth]{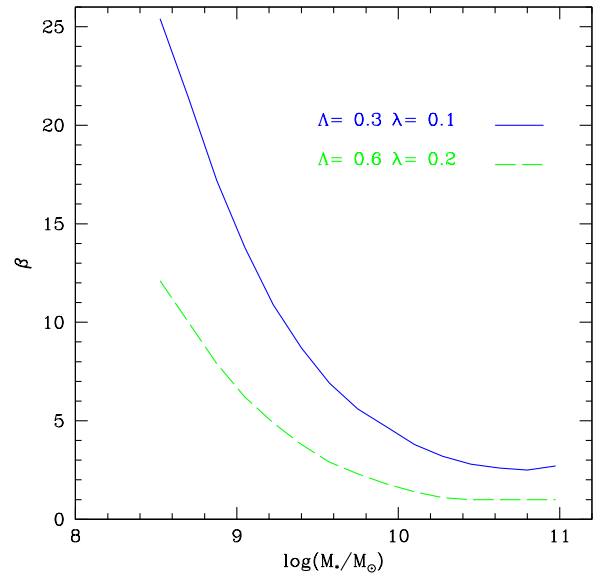}
\caption{The best values for $\beta$ in a model with both infall and outflow. The blue solid line is referred to a system with $\Lambda=0.1$ and $\lambda=0.3$, whereas the results for  $\Lambda=0.2$ and $\lambda=0.6$  are indicated with the long dashed green line.} 
\label{alfarecchi}
\end{figure}

We also study the case in which $\Lambda>\lambda$, with this set of parameters: $(\Lambda,\lambda)=(0.3,0.1);(0.6,0.2),(3,1)$. In this case, the solution is  defined only for values of $\mu$ such that:         
\begin{equation}
\mu > {{\Lambda - \lambda - 1} \over {\Lambda - \lambda}},
\label{eq:mumin2}
\end{equation}
as in the case of the model with outflow and infall.  Using the set
(3,1), $\mu$ must be larger than 0.5. The observed average $\mu$ spans
the range between 0.16 and 0.47, then the (3,1) set of parameters
should be discarded.  Both the first and the second model can
reproduce the MZ relation, but the set (0.3,0.1) requires values of
$\beta$ too high and unrealistic in comparison with the wind parameter
$\lambda$, as shown in Fig. \ref{alfarecchi}, and therefore it should
also be rejected.

\begin{table*}[htp]

\caption{Permitted values of the parameters for reproducing the MZ relation. }
\scriptsize

\label{edmund}
\begin{center}
\begin{tabular}{c|ccccc}
  \hline
\noalign{\smallskip}


\\
 Our Models &$\Lambda$&$\lambda$ & $\beta$&$y_0$   \\

  \\
\noalign{\smallskip}

\hline
\noalign{\smallskip}

Only outflow& / &  $  0.67\leq \lambda\leq5.55    $ &/&0.01\\
\noalign{\smallskip}
\hline
\noalign{\smallskip}
Only infall & see Fig. 6  &  $  /  $ &/&0.01\\
\noalign{\smallskip}
\hline
\noalign{\smallskip}
Outflow + infall (varying both  $\Lambda$ and $ \lambda$) & $  0.10 \leq \Lambda \leq2.51$ & $  0.11 \leq \lambda \leq 5.40$ &/&0.01\\
\noalign{\smallskip}
 \hline
\noalign{\smallskip}

Outflow + infall (fixed $\Lambda$  and  varying$ \lambda$) & $  0 < \Lambda \leq 0.70$ &   if $\Lambda=0.70$; $0.16 \leq \lambda \leq 0.47$  &/&0.01\\

& &    if $\Lambda=0.01$; $0.66 \leq \lambda \leq 5.54$      & /&0.01\\
\noalign{\smallskip}

 \hline
\noalign{\smallskip}
 Enriched Outflow + infall with $\Lambda<\lambda$ ( fixed $\Lambda=0.1$) & 0.1   & $0.10<\lambda \leq 0.60$ &  if $\lambda = 0.60 $; $1.0\leq \beta \leq 5.0  $ &0.01\\
& &  &  if $\lambda=0.11$; \, $3.3 \leq \beta \leq 24.4$      &0.01\\
\noalign{\smallskip}

 \hline
\noalign{\smallskip}
 Enriched Outflow + infall with $\Lambda>\lambda$ ( fixed $\Lambda=0.30$ and $\lambda=0.10$ ) & 0.3   & 0.1 &  $  2.5 \leq \beta \leq 25.4  $ &0.01\\
\noalign{\smallskip}

 \hline
\noalign{\smallskip}

 Enriched Outflow + infall with $\Lambda>\lambda$ ( fixed $\Lambda=0.6$ and $\lambda=0.2$ ) & 0.6   & 0.2 &  $  1.0 \leq \beta \leq12.1  $ &0.01\\
\noalign{\smallskip}

 \hline
\noalign{\smallskip}

 Closed box+ variable IMF & /  & / &    / &    $  0.004 \leq  y_0\leq  0.007   $\\
\noalign{\smallskip}

\hline
\end{tabular}
\end{center}
\end{table*}

 In Table \ref{edmund} we summarize the permitted values of the parameters for reproducing the MZ relation for all the cases studied in this work.

\section{Conclusions}

In this paper we have explored different solutions for the MZ relation observed in SDSS 
galaxies. We adopted simple analytical models including infall, outflow and differential outflow besides the classical closed-box case. All the solutions of the analytical models contain the yield per stellar generation plus several parameters describing the different physical processes considered.
We started by deriving the gas mass for the sample of the studied galaxies with known stellar galactic masses and star formation rates, by inverting the K98 law. Then, by means of these gas masses we derived the metallicity of each galaxy through the analytical models.
At this point, we tried to vary the  model parameters  corresponding to the different physical processes to see whether the observed MZ could be reproduced.
As it is well known, in order to obtain that metallicity increasing with galactic mass, the effective yield should decrease with decreasing mass. The effective yield can decrease because the outflow or infall become more important  with decreasing mass or because the IMF becomes steeper. Some of these cases have been already studied by other authors but never all together, and here we add a new case with differential outflow with wind intensity increasing with decreasing mass. By differential outflow we mean a galactic wind where metals are lost preferentially. Then, we tested whether the required parameter variations were realistic in order to choose among the different solutions.

Our conclusions can be summarized as follows:
\begin{itemize}

\item 
The observed local MZ relation can be perfectly reproduced
 under the assumption of a closed-box and an effective yield varying
 from galaxy to galaxy. This means that either the nucleosynthesis is
 different or, more likely, that the IMF is different from galaxy to
 galaxy.  The required yield variation is rather small and lies in the
 range 0.004-0.007, corresponding to a variation of the slope of the
 IMF (one-slope IMF defined in the mass range 0.1-100$M_{\odot}$) from
 $\sim$1.35 to 1.5 going from more to less massive galaxies. However,
 the variation of the IMF always implies variations in other physical
 quantities besides the O abundance and therefore, before accepting
 this solution, it would be necessary to test other observables such
 as colors, the color-magnitude diagrams and the M/L ratios, if
 available.  Besides that, if we change the IMF we should in principle
 change also the conversion factor from SFR to $\mu$ in the inversion
 of the Kennicutt law, which in our work has been calculated by
 adopting a Salpeter IMF in the mass range 0.1-100 $M_{\odot}$ (see
 Sect. 3).

\item The observed MZ relation can be very well reproduced by a
  constant yield but a variable efficiency of the outflow, increasing
  from more to less massive galaxies.   This is not a new
    conclusion since it has been discussed before by several authors
    (Garnett 2000, Tremonti et al. 2004, Edmunds 2005). However, here
    we reached this conclusion by considering the most recent
    observational data.  In particular, the wind parameter $\lambda$,
  which roughly represents the ratio between the wind rate and the
  SFR, should vary from 1 to 5.5, going from large to small galaxies,
  which is quite a reasonable range supported also by galactic wind
  observations (Martin et al. 2002). We have also found that outflows
  must play an important role especially for galaxies with stellar
  masses $<10^{10} M_{\odot}$. This is consistent with other results
  on the efficiency of outflows as a function of galactic mass, which
  show that dwarf galaxies must have ejected a fraction of their
  present baryonic mass larger than massive galaxies (e.g. Gibson \&
  Matteucci 1997; Calura et al. 2008).
\item The local MZ relation cannot be reproduced by a variable infall
  efficiency no wind and constant IMF, so this solution should be
  rejected. This does not mean, however, that the infall is not
  important for the considered galaxies.Infall can be present but it
  is not the cause of the MZ relation. Moreover, an infall rate
  increasing with decreasing galactic mass it is not so easy to
  explain under physical basis.

\item Models with variable infall and outflow rates can, in fact, well 
reproduce the observed local MZ relation. In this case, the outflow rate is
generally larger than the infall rate.

\item
 Differential galactic winds, where mostly metals are lost from the
 galaxy can also very well reproduce the observed MZ relation,
 provided that $\lambda>\Lambda$. In this case, the metal ejection
 efficiency $\beta$ is always larger in smaller galaxies.
On the other hand, when $\Lambda>\lambda$ not all the values of
$\Lambda$ and $\lambda$ are acceptable. Considering only the
acceptable values for these parameters, we found that $\beta$ values,
able to reproduce the MZ relation, are generally higher than in the
case with $\Lambda<\lambda$.

 The assumption of a larger $\beta$ in low-mass galaxies simply means
 that these galaxies are able to expel large fractions, relative to
 their total mass, of newly synthesized metals but only tiny fractions
 of pristine ISM, in agreement with many hydrodynamical studies of
 star forming dwarf galaxies (e.g. MacLow \& Ferrara 1999, D'Ercole \&
 Brighenti 1999, Recchi et al. 2001).

\item We did not explore the case in which the efficiency of star
  formation increases with galactic mass, since the SFR does not
  appear in the solutions of the analytical models. Moreover, such a
  solution has already been explored in detail by Calura et
  al. (2009), who showed that this assumption well reproduces the MZ
  relations both at low and high redshift.  In addition, this
  variation of the star formation efficiency produces a downsizing
  effect in the star formation often invoked to explain the properties
  of ellipticals (Pipino \& Matteucci, 2004).

\item In conclusion, on the basis of this paper and of Calura et al.'s
  (2009) previous results, the most plausible solution is that the MZ
  relation is created by a variable star formation efficiency from
  galaxy to galaxy coupled to galactic winds (preferentially
  metal-enhanced) becoming more and more important in low mass
  galaxies.  But as stressed by Edmunds (2005), explaining the
mass-metallicity relation by a systematic increase in the star
formation efficiency with galactic mass poses the problem of
understanding the physical mechanism behind it. Several suggestions
exist in the literature but the real physical origin of this
correlation is not known. A possibility is that the higher the
pressure in the ISM then the higher is the star formation efficiency,
and the more massive galaxies have higher ISM pressure because of
their deeper potential wells (see Elmegreen \& Efremov 1997; Harfst et
al. 2006).

\end{itemize}

\acknowledgements{We thank the referee, M. Edmunds, for the enlightening suggestions.  We acknowledge financial support from  MIUR (Italian Ministry of Research, contract 
PRIN2007 Prot.2007JJC53X-001) and from Italian Space Agency contract ASI-INAF I/016/07/0.}

\end{document}